\documentclass[prb,twocolumn,showpacs,preprintnumbers,amsmath,amssymb]{revtex4}
\usepackage{graphicx}
\usepackage{dcolumn}
\usepackage{bm}
\usepackage{textcomp}

\begin{document}

\preprint{APS/123-QED}

\title{Nucleation of Ge quantum dots on the Si(001) surface}

\author{Larisa V. Arapkina}

\author{Vladimir A. Yuryev}
\email{vyuryev@kapella.gpi.ru}

\affiliation{A.\,M.\,Prokhorov General Physics Institute of the Russian Academy of Sciences,\\ 38 Vavilov Street, Moscow, 119991, Russia}

\date{\today}%

\begin{abstract}
A direct observation of nucleation of Ge hut clusters formed by ultrahigh vacuum molecular beam epitaxy is reported. 
The nuclei  of the pyramidal and wedge-like clusters have been observed on the wetting layer blocks and found to have different structures. The  growth of the clusters of both species goes on following different scenarios:  Formation of the second atomic layer of the wedge-like cluster results in rearrangement of its first layer. Its ridge structure does not replicate the structure of the nucleus. The  pyramidal cluster grows without 
phase transitions. The structure of its vertex copies the structure of the nucleus. 
The wedge-like clusters contain point defects on the triangular faces and have preferential directions of growth along the ridges.

\end{abstract}

\pacs{68.37.Ef, 81.07.Ta}
\maketitle

\begin{figure}[b]
\includegraphics[scale=.6]{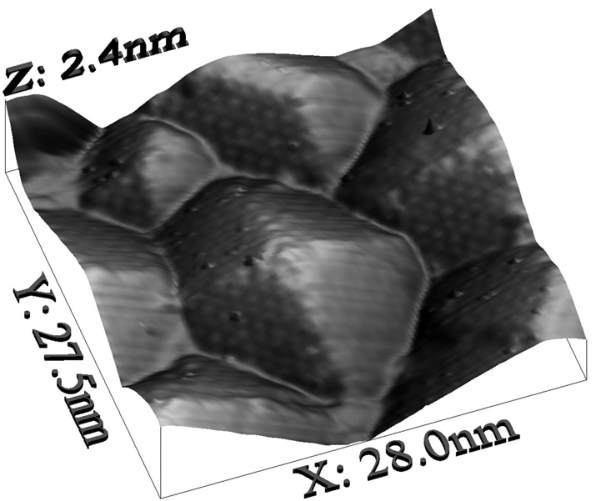}(a)
\includegraphics[scale=.65]{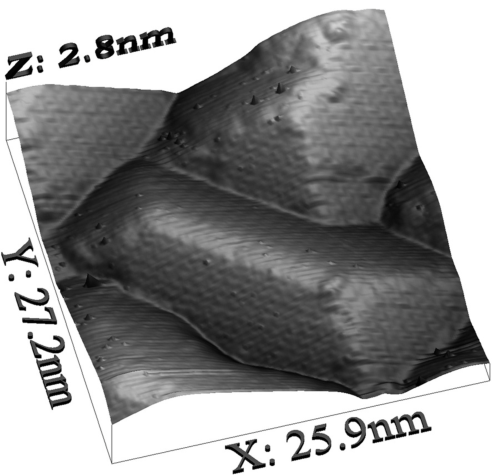}(b)
\includegraphics[scale=.7]{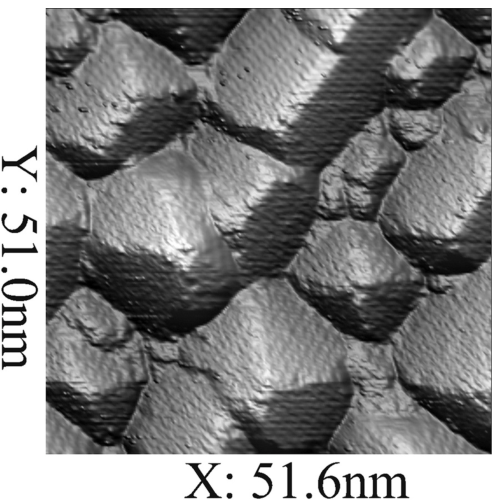}(c)
\includegraphics[scale=.7]{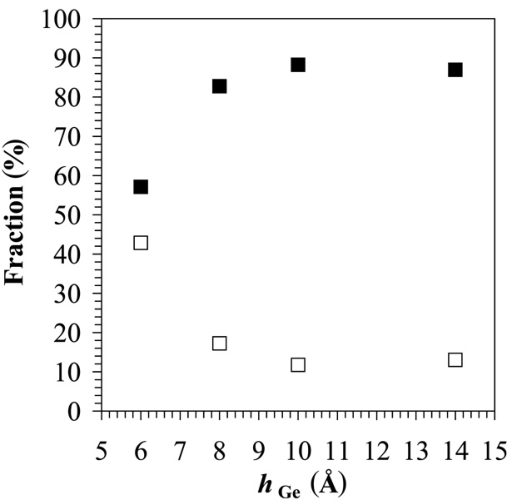}(d)
\caption{\label{fig:example}STM images of Ge pyramidal (a) and wedge-like (b) clusters, Ge QD dense array ($h_{\rm Ge}=10$\,\r{A}) on the Si(001) surface (c), and a fraction of wedges ($\blacksquare$) and pyramids ($\square$) in the arrays (d)  {\it vs} Ge coverage ($T_{\rm gr}=360^{\circ}$C).}
\end{figure}

Arrays of densely packed  self-assembled Ge quantum dots (QD) on the Si(001) surface (Fig.~\ref{fig:example}) \cite{Smagina,classification} due to the phenomenon of quantum confinement of carriers are currently considered as a basis for development  of prospective devices of photoelectronics \cite{Wang-properties,
Pchel_Review}. 
Extensive investigations carried out for the last two decades 
(see, e.g., Refs.~\onlinecite{Mo,Chem_Rev,Nucleation,Kastner,Island_growth,Fujikawa,Facet-105,Ge_QD_crystal})
resulted in the
technological achievements of the recent years that enabled the controllable formation of Ge QD arrays with the desired cluster densities (up to $10^{12}$~cm$^{-2}$, Refs.~\onlinecite{Smagina,classification}). 
However, the problems of uniformity of cluster types in the arrays and the dispersion of cluster  sizes are still  far from solution. That is why the intensive investigations of the cluster morphology and growth process  with view of reproducible formation of uniform and defectless QD arrays are strongly required. This is an issue of special importance for the ordered QD arrays \cite{Ge_QD_crystal} taking into account extremely exacting restrictions imposed on the uniformity by the aim of development of such arrays. Non-uniform ordered array composed of clusters of different types and sizes would not operate as 3D crystal of artificial atoms, or even would not reproduce regularity in successive QD layers  if containing defects such as large and extended clusters or depleted regions \cite{defects_ICDS-25}.

\begin{figure*}
\includegraphics[scale=1]{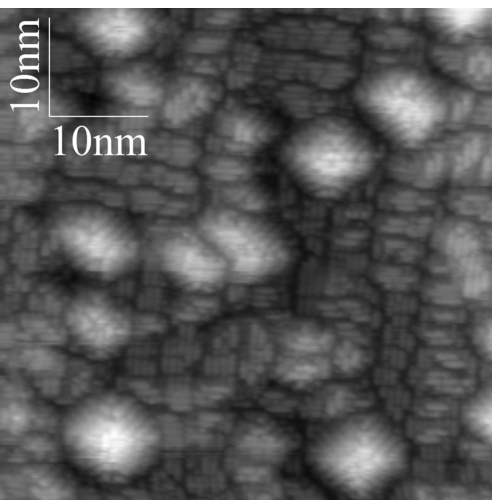}(a)
\includegraphics[scale=1]{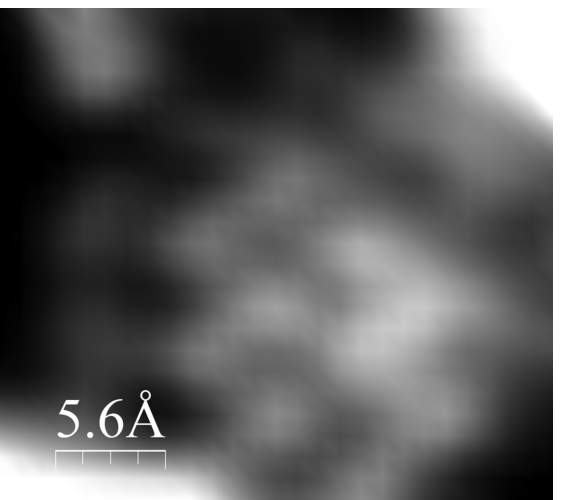}(b)
\includegraphics[scale=1]{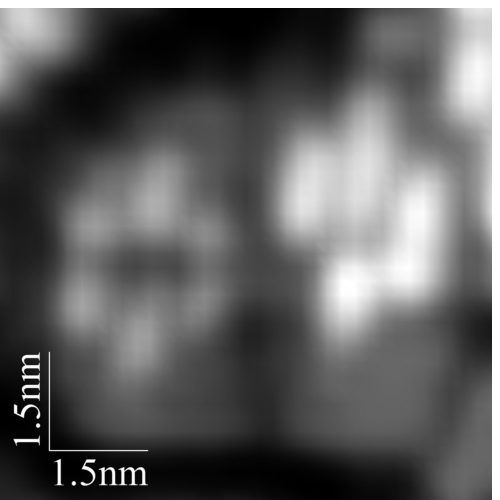}(c)
\caption{\label{fig:arrays}STM empty state image of Ge QD array ($h_{\rm Ge}=6$~\r{A},  $T_{\rm gr}=360^{\circ}$C) on the Si(001) surface (a);  $p(2\times 2)$ structure within the WL block, upper Ge atoms of the tilted dimers are resolved in the rows (b); pyramid (left) and wedge nuclei (1\,ML) on the neighboring WL blocks (c), both nuclei reconstruct the WL surface, a nucleus never  exceeds the bounds of a single WL block. }
\end{figure*}

\begin{figure}
\includegraphics[scale=.75]{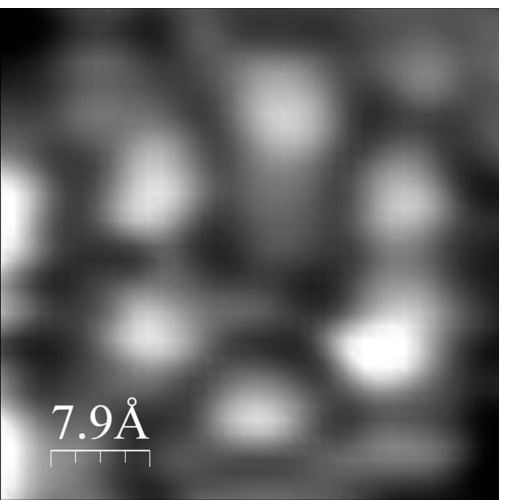}(a) 
\includegraphics[scale=.75]{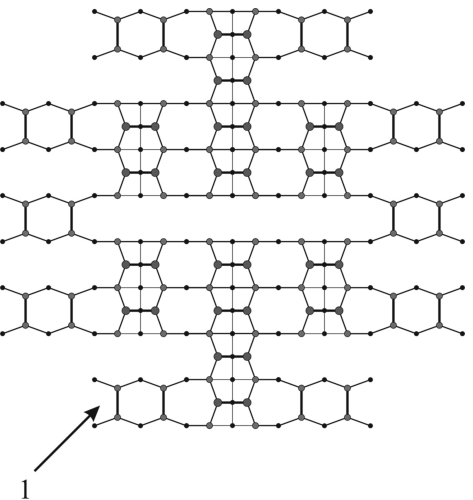}(b) 
\includegraphics[scale=.85]{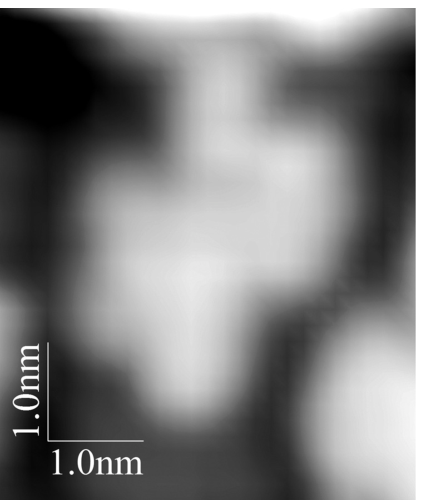}(c)~~~~~\,
\includegraphics[scale=.8]{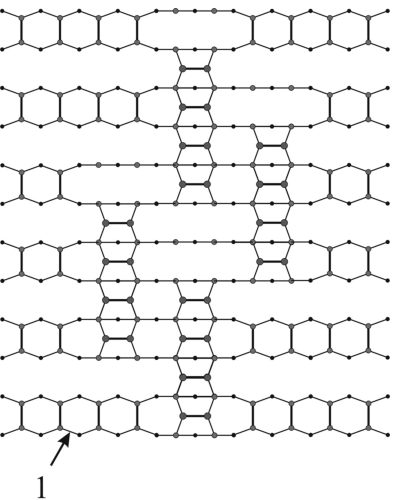}(d) 
\caption{\label{fig:nuclei}Nuclei of Ge hut clusters: STM empty state images (a,\,c) and atomic structures (b,\,d) of the pyramid (a,\,b)   
 and  wedge (c,\,d) nuclei, 1 is WL.}
\end{figure}

Recently we showed  that the  \{105\} faceted clusters usually referred to as hut clusters \cite{Mo} are subdivided into two main {\it morphologically different} species---pyramids and wedges (Fig.~\ref{fig:example}) \cite{classification}. In the literature, both species of hut clusters are traditionally considered as structurally identical and genetically connected types \cite{Mo,Facet-105}.  Explanations of transitions from square shaped to elongated islands (from pyramids to wedges in our terminology) are discussed \cite{Nucleation,Kastner,Island_growth} although no clear observations of such phenomenon have been described anywhere. Different models from simple coalescence of neighboring square shaped clusters \cite{Nucleation} to more sophisticated kinetic  model of growth \cite{Kastner} have been brought forward which are in satisfactory agreement with observations. We found that at moderate growth temperatures the densities of clusters of both species are equal at the initial stage of the array formation (Fig.~\ref{fig:example}(d)). Then, as the Ge coverage is increased, the wedges become dominating in the arrays whereas the pyramids exponentially rapidly disappear \cite{classification,endnote_1}. 
Lately we investigated by STM the structure of the $\{105\}$ cluster facets together with the structure of apexes (ridges  and vertices) of the clusters and built structural models of both species of huts \cite{atomic_structure}. We found the structure of the ridges of the wedge-like clusters to be different from the structure of the vertices of the pyramidal ones, therefore a wedge-like cluster cannot arise from a pyramidal one and vice versa \cite{classification,atomic_structure}. Transitions between the shapes of the hut clusters are prohibited \cite{endnote_3}. One can find additional evidences of the above strong statement investigating the cluster nucleation and the initial  stage of its growth by {\it in situ} STM with high enough resolution.

At present, nucleation of Ge clusters on the Si(001) surface is still very little-studied. Probably only two direct observations of this phenomenon were reported by Goldfarb {\it et~al.}~\cite{Nucleation,Goldfarb_JVST-A} and Vailionis {\it et~al.}~\cite{Vailionis}. Those comprehensive {\it in situ} STM studies explored  gas-source-molecular-beam-epitaxy (GS-MBE) growth of Ge on Si(001) in the atmosphere of GeH$_4$ \cite{Nucleation,Goldfarb_JVST-A} or Ge$_2$H$_6$ \cite{Vailionis}. The chemistry of GS-MBE is obviously strongly different from that of ultrahigh vacuum (UHV) MBE which is usually employed for Ge deposition on Si substrates \cite{Pchel_Review}. Unfortunately, experimental and especially direct high resolution UHV STM investigations of Ge cluster nucleation and early stages of the cluster growth on Si(001) by UHV MBE have not been described in the literature thus far. No data are available on the morphology of nuclei and the beginning of cluster growth. Now we shall try to fill up this gap.

\begin{figure*}
\includegraphics[scale=1]{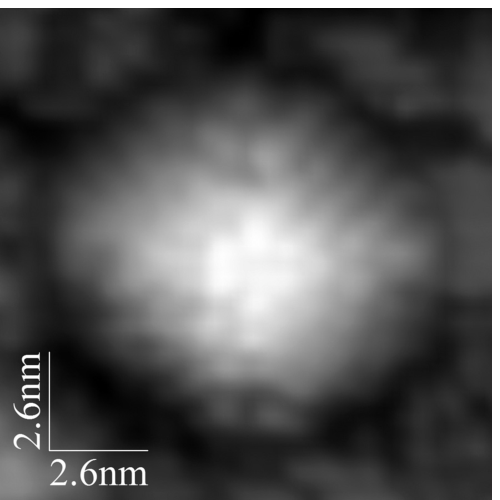}(a)~~~~\,
\includegraphics[scale=1]{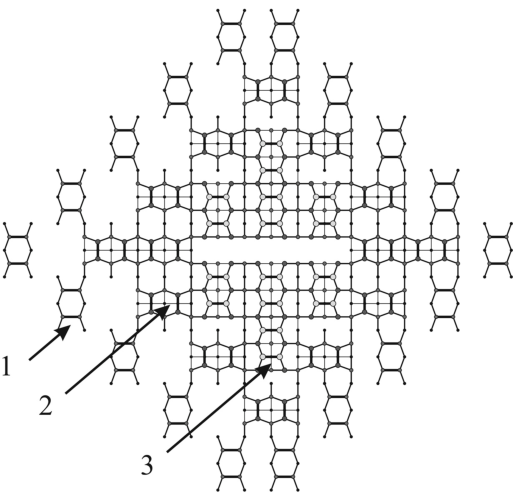}(b)~~~~
\includegraphics[scale=1]{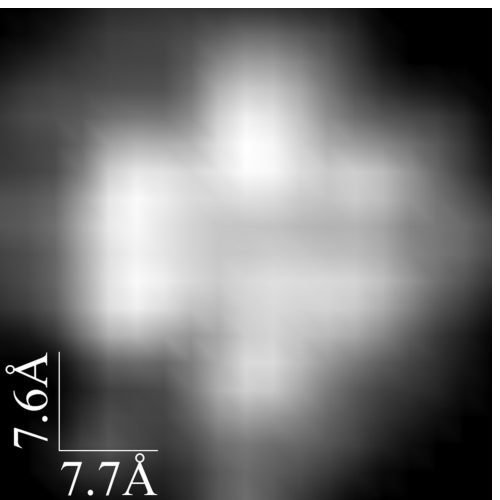}(c)
\\
\includegraphics[scale=1]{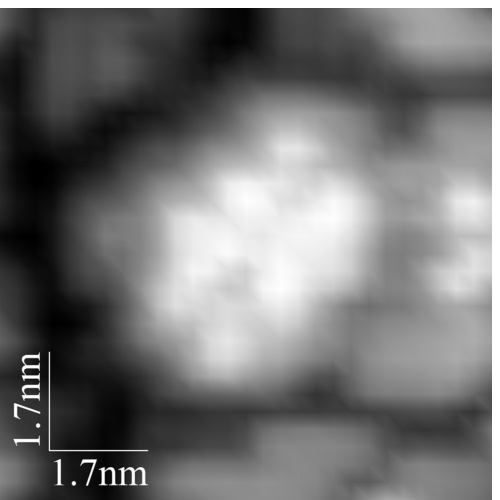}(d)
\includegraphics[scale=1]{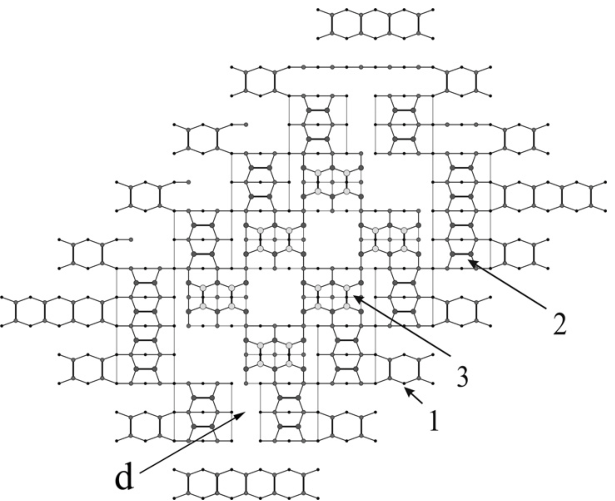}(e)
\includegraphics[scale=1]{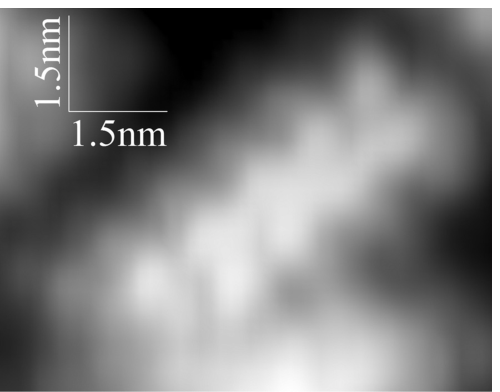}(f)
\caption{\label{fig:dots} STM empty state micrograph (a) of the 5-ML Ge pyramid ($h_{\rm Ge}=6$~\r{A}, $T_{\rm gr}=360^{\circ}$C), a top view of the pyramidal QD (b) and   contrasted image of its vertex (c); STM empty state topographs ($h_{\rm Ge}=6$~\r{A}, $T_{\rm gr}=360^{\circ}$C)   of the 2-ML Ge wedge-like cluster (d), a top view of the wedge-like QD (e) and an empty state image of the ridge of the 3-ML Ge wedge-like cluster (f); 1, 2 and 3 designate WL, the first and the second layers of QD respectively, d marks a defect arisen because of one translation uncertainty of the left dimer pair position. }
\end{figure*}

In this article, we investigate the nucleation and very beginning of growth of Ge hut clusters composing dense QD arrays formed by UHV MBE at moderate temperatures. 
The atomic structure of cluster nuclei as well as the structures of very little clusters---as small as a few monolayers (ML) high over the wetting layer (WL)---are the issues of this study \cite{sabelnik}.

The results reported in the article evidence that there are two different types of nuclei on Ge wetting layer which evolve in the process of Ge deposition to pyramidal and wedge-like hut clusters. It might seem that solid proofs of this statement can be only obtained from STM
measurements during growth \cite{Nucleation, Kastner}. Unfortunately, such  experiment is hardly possible now. STM operating at the growth temperatures cannot assure atomic resolution which is necessary to reveal an atomic structure of clusters and smaller objects  on WL. We have made a different experiment. Having assumed that nuclei emerge on WL as combinations of dimer pairs and/or longer chains of dimers in epitaxial configuration \cite{epinucleation} and correspond to the known structure of apexes specific for each hut species \cite{classification, atomic_structure}
we have investigated WL patches, 1\,ML high formations on them and clusters of different heights (number of steps) over WL. This approach exactly simulates the above experiment ensuring the required high resolution. As a result, we succeeded to select two types of formations different in symmetry and satisfying the above requirements, which first appear  at a coverage of $\sim 5$~\r{A} ($T_{\rm gr}=360^{\circ}$C) and then arise on  WL during the array growth. We have interpreted them as hut nuclei, despite their sizes  are much less than those predicted by the first principle calculations \cite{hut_stability}, and
traced their evolution to 
huts \cite{endnote_7}.

\begin{figure}
\includegraphics[scale=.85]{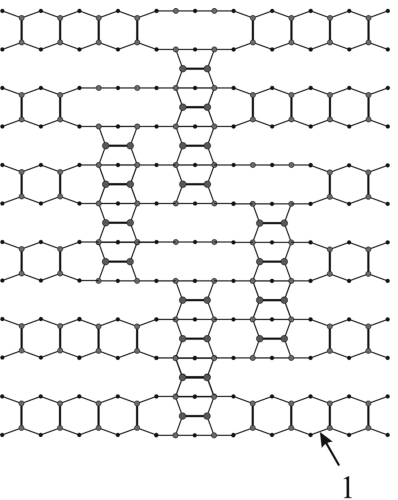}(a)
\includegraphics[scale=.85]{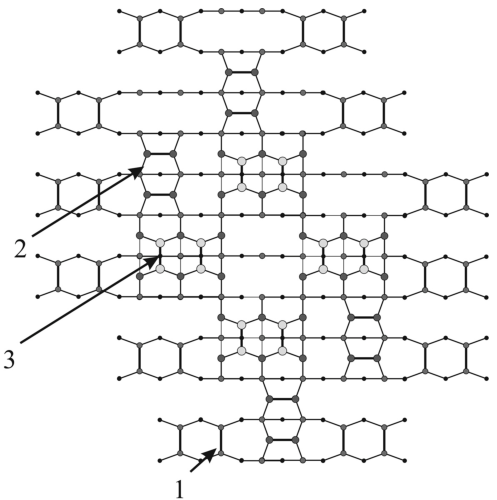}(b)
\caption{\label{fig:phase_transition}
Rearrangement of the first layer (a) of  a forming   wedge during addition of dimer pairs of the second layer  (b); labels are the same  as in Fig.~\ref{fig:dots}.}
\end{figure}


The experiments were carried out using an ultra high vacuum instrument consisting of the UHV MBE chamber coupled with high resolution STM which enables the sample study at any stage of processing sequentially investigating the surface and giving additional treatments to the specimen; the samples never leave UHV ambient during experiments. 
Silicon substrates ($p$-type, $\rho = 12~\Omega$\,cm) were completely deoxidized as a result of short annealing at the temperature of $\sim 925^\circ$C \cite{our_Si(001)_en}. Germanium was deposited directly on the atomically clean Si(001) surface from the source with the electron beam evaporation  \cite{classification}.
The rate of Ge deposition was  $\sim 0.1$~\r{A}/s and the Ge coverage  ($h_{\rm Ge}$) \cite{endnote_4} was 
varied from 3 to 14~\r{A}.  
The substrate temperature $T_{\rm gr}$ was $360^\circ$C during Ge deposition. The rate of the sample cooling down to the room temperature was  $\sim 0.4^\circ$C/s after the deposition. 
The temperature was monitored with tungsten-rhenium thermocouple mounted in vacuum near the rear side of the samples and {\it in situ} graduated beforehand against the IMPAC~IS\,12-Si pyrometer which measured the sample temperature through the chamber window.  
Specimens were scanned at room temperature in the constant tunneling current ($I_{\rm t}$) mode. The STM tip was zero-biased while a sample was positively or negatively biased ($U_{\rm s}$). The details of the sample preparation as well as the experimental techniques can be found elsewhere \cite{classification, our_Si(001)_en, WSxM}.


Fig.~\ref{fig:arrays}(a) presents an  STM image of an array of small Ge clusters grown at $T_{\rm gr}=360^{\circ}$C and $h_{\rm Ge}=6$~\r{A}.  WL is seen to have a block ($M \times N$ patched) structure. The blocks are usually $p(2\times 2)$ reconstructed (Fig.~\ref{fig:arrays}(b)) \cite{endnote_5}. We suppose that the process of the cluster nucleation consists in formation of new structures on the WL blocks. These 1\,ML high structures are well resolved in Fig.~\ref{fig:arrays}(c) on the neighboring WL blocks: The left feature is assumed to be a nucleus of the pyramid whereas the right one is considered as a nucleus of the wedge-like cluster. A good few of such structures are observed in the long shot of the array (Fig.~\ref{fig:arrays}(a)). STM images of the nuclei and their schematic plots are given in Fig.~\ref{fig:nuclei}. The further growth of the clusters is shown in Fig.~\ref{fig:dots}. Fig.~\ref{fig:dots}(a) presents an STM image of the 5\,ML high pyramid. It is commonly adopted that the hut clusters grow by successive filling the (001) terraces of the $\{105\}$ faces by the dimer rows \cite{Kastner}. 
 A schematic plot of the 2-ML pyramid  based on this assumption (Fig.~\ref{fig:dots}(b)) demonstrates its atomic structure (even number of layers is shown in both (a) and (b) pictures, so the diagram  reproduces the entire structure of the dot except for its height). It is seen comparing Figs.~\ref{fig:nuclei}(a) and \ref{fig:dots}(c) that the vertex repeats the structure of the nucleus drown in Fig.~\ref{fig:nuclei}(b)\cite{fig1}. 
The characteristic distances exactly match.
The $<$100$>$ direction of the base sides is predetermined by the nucleus structure, thus the pyramids  grow  without phase transition when the second and subsequent layers are added. Only nucleus-like structures of their apexes are rotated $90^{\circ}$ with respect to the rows on previous terraces to form the correct epitaxial configuration when the heights are increased by 1\,ML, but this rotation does not violate the symmetry of  the previous layers of the cluster. 

\begin{figure}[t]
\includegraphics[scale=1.2]{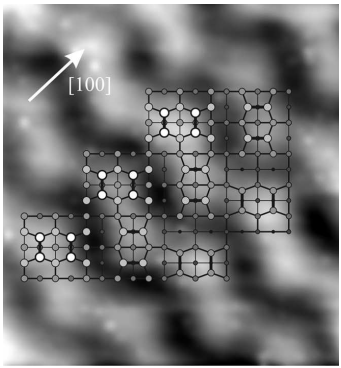}
\caption{\label{fig:face}
Schematic  drawing of the $\{105\}$ facet  superimposed on its STM  image  ($4.3\times 4.4$~nm, $U_{\rm s}=+3.0$~V, $I_{\rm t}=100$~pA), the cluster base side is parallel to the [100] direction, the steps rise from the lower right to the upper left corner.}
\end{figure}

A different scenario of growth  of the wedge-like clusters have been observed. Figs.~\ref{fig:dots}(d,\,e) show an image and a schematic diagram of the 2-ML wedge-like cluster. The ridge structure is seen to be different from the nucleus structure presented in Figs.~\ref{fig:nuclei}(c,\,d). The structure of the ridge is well resolved in the image of the 3-ML cluster (Fig.~\ref{fig:dots}(f)) filtered to contrast the uppermost layer of atoms.  In this image, the dimer pairs of the ridge are $90^{\circ}$ rotated compared to the 2-ML wedge that is in full agreement with the proposed atomic model\cite{ridge}. This structure of the wedge-like cluster arise due to rearrangement of  rows of the first layer in the process of the second layer formation (Fig.~\ref{fig:phase_transition}).  The phase transition in the first layer  generates the base  with all sides directed along the $<$100$>$ axes which is necessary to give rise to the $\{105\}$ faceted cluster (it is seen from Figs.~\ref{fig:nuclei}(c,\,d) that only one pair of sides of a wedge nucleus runs along the  $<$100$>$ direction). After the transition, the elongation of the elementary structure is possible only along a single axis which is determined by the symmetry and clearly seen when comparing  Figs.~\ref{fig:dots}(e) and \ref{fig:phase_transition}(b) (along the arrows in Fig.~\ref{fig:phase_transition}(b)).
This preferential growth direction determines the rapid growth on the triangular facets (short edges). 
The growth on these facets does not change the orientation of the dimer pairs forming the ridge. It is obviously also that it cannot increase the cluster height but only its length. The increase of the cluster height is governed by the completion of the trapezoidal facet \cite{facet_growth}. The latter process is accompanied by the change of direction of the  dimer pairs on the ridge when the apex terrace is completed. Note that the phenomenon of the wedge height limitation described in Ref.~\onlinecite{classification} differs from the process of its length self limitation. The former is mainly controlled by the growth temperature and the later is governed by either the area of the trapezoidal faces  or the number and/or sizes of the WL blocks covered by the elongating cluster, as well as the competition of the processes of the in-height and longitudinal growth. In general, the cause of the wedge elongation is still unclear now.

It is necessary to remark here that the nuclei are always observed to arise on sufficiently large WL patches. There must be enough room for a nucleus on a single patch. A nucleus cannot be housed on more than one patch. So, cluster nucleation is impossible on little (too narrow or short) patches (Fig.~\ref{fig:arrays}(a)).

It should be noted also that according to the proposed model the wedge-like clusters always contain point defects on the triangular (short) facets. The defects are located in the upper corners of the facets  and caused by uncertainty of one translation in the position a dimer pair  which forms the penultimate terrace of the triangular facet (Figs.~\ref{fig:dots}(d--f)). The predicted presence of these defects removes the degeneracy of the facets and hence 
an issue of the symmetry violation which occur if the pyramid-to-wedge transition is assumed (this issue was discussed in detail in Ref.~\onlinecite{classification}). These defects are absent on the facets of the pyramidal huts. 
Their triangular facets  are degenerate. Therefore, as it follows from our model, the trapezoidal and triangular facets of the wedge 
are not degenerate with respect to one another even at very beginning of cluster growth. The wedges can easily elongate by growing on the 
triangular facets faster than on trapezoidal ones, whereas pyramids, having degenerate facets, cannot elongate and grow 
only in height outrunning wedges. This explains greater heights of pyramids \cite{classification}.

The proposed models being applied to draw the clusters by filling terrace by terrace (like it is done in Fig.~\ref{fig:dots}) allowed us to 
deduce a model of the $\{105\}$ facets. This model resulting from the above simple crystallographic consideration corresponds to the PD \cite{Mo} (paired dimers)  rather than more recent RS (rebonded step) model \cite{Fujikawa,Facet-105} which is now believed to improve the previous PD model by Mo {\it et al}. Being superposed with the empty state STM image of the cluster  $\{105\}$ facet it demonstrates an excellent agreement with the experiment (Fig.~\ref{fig:face}). Dangling bonds of the derived in such a way $\{105\}$-PD facets in reality
 may stimulate Ge atom addition  and cluster growth.
Less stability of the $\{105\}$-PD  facets compared to the Ge(105)/Si(105)-RS plane may cause fast completion of hut terraces during epitaxy.

It should be noticed also that, as it follows from the reported  models, the growth of the wedge second layer  requires reconstruction of the buried previous layer.
This phenomenon has been discussed theoretically before as ``critical
epinucleation'' on reconstructed surface \cite{epinucleation}. In particular, the atomic
models drawn in Figs.~\ref{fig:dots}(e) and~\ref{fig:phase_transition} show the ad-dimer rows
un-reconstructing  the surface layer that can only happen beyond
a critical number of ad-dimers defined as ``epinucleus''. So, the presented data
could be one of the first experimental evidence of the
epinucleus  \cite{thanks}. The critical epinucleation appears to be a basic phenomenon for hut  formation on $(M\times N)$.

 In conclusion, 
we have reported the direct observation of nucleation of Ge hut clusters formed by UHV MBE on the Si surface.
The nuclei  of the pyramidal and wedge-like clusters have been observed on the wetting layer $(M\times N)$ patches and found to have different structures. The atomic models of nuclei of both species of the hut clusters have been built as well as the models of the clusters at the early stage of growth.
The  growth of the clusters of each species has been demonstrated to follow generic scenarios. The  formation of the second atomic layer of the wedge-like cluster results in rearrangement of its first layer. Its ridge structure does not repeat the structure of the nucleus. 
The  pyramidal cluster grows without phase transitions. The structure of its vertex copies the structure of the nucleus. The cluster of one species cannot turn into the cluster of the other species.
The wedge-like clusters contain point defects in the upper corners of the triangular faces and have preferential directions of growth along the ridges.
The derived structure of the $\{105\}$ facet corresponds to the PD  model. The critical epinucleation phenomenon may be responsible for  hut  formation on $(M\times N)$ patched WL.









\begin{thebibliography}{99}

\bibitem{Smagina}
J. V. Smagina, 
V. A. Zinovyev, A. V. Nenashev, A. V. Dvurechenski\u{i}, V. A. Armbrister, and S. A. Teys,
JETP {\bf 106}, 517 (2008);

\bibitem{classification}
L. V. Arapkina and V. A. Yuryev, Physics-Uspekhi {\bf 53}, 279 (2010).

\bibitem{Wang-properties}
K. L. Wang, 
S. Tong, and H. J. Kim, 
Mater. Sci. Semicond. Proc. {\bf 8}, 389 (2005);
K. L. Wang, 
D. Cha, J. Liu, and C. Chen,
Proc. IEEE {\bf 95}, 1866 (2007).

\bibitem{Pchel_Review}
O. P. Pchelyakov, 
Yu. B. Bolkhovitjanov, A. V. Dvurechenski\u{i}, L. V. Sokolov, A. I. Nikiforov, A. I. Yakimov, and B. Voigtl{\"{a}}nder,
Semicond. {\bf 34}, 1229 (2000).

\bibitem{Mo}
Y.-W. Mo, D. E. Savage, B. S. Swartzentruber, and M. G. Lagally, 
Phys. Rev. Lett. {\bf 65}, 1020 (1990).

\bibitem{Chem_Rev}
F. Liu, F. Wu, and M. G. Lagally, Chem. Rev. {\bf 97}, 1045 (1997).

\bibitem{Nucleation}
I. Goldfarb, P. T. Hayden, J. H. G. Owen, and G. A. D. Briggs, 
Phys. Rev. Lett. {\bf 78}, 3959 (1997).

\bibitem{Kastner}
M. K{\"{a}}stner and B. Voigtl{\"{a}}nder, 
Phys. Rev. Lett. {\bf 82}, 2745 (1999).



\bibitem{Island_growth}
D. E. Jesson, G. Chen, K. M. Chen, and S. J. Pennycook, 
Phys. Rev. Lett. {\bf 80}, 5156 (1998)


\bibitem{Fujikawa}
Y. Fujikawa, K. Akiyama, T. Nagao, T. Sakura, M.~G.~Lagally, T. Hashimoto, Y. Morikawa, and K. Terakura,
Phys. Rev. Lett. {\bf 88}, 176101 (2002).

\bibitem{Facet-105}
P. Raiteri , D.~B. Migas, L. Miglio, A. Rastelli, and H. von K{\"{a}}nel,
Phys. Rev. Lett. {\bf 88}, 256103 (2002).

\bibitem{Ge_QD_crystal}
D. Gr{\"{u}}tzmacher, T. Fromherz, C. Dais, J. Stangl, E. M{\"{u}}ller, Y. Ekinc, H. H. Solak, H. Sigg, R. T. Lechner, E. Wintersberger, S. Birner, V. Hol{{y}}, and G. Bauer,
Nano Lett. {\bf 10}, 3150 (2007).


\bibitem{defects_ICDS-25}
V. A. Yuryev  and L. V. Arapkina, Physica B {\bf 404}, 4719 (2009). 

\bibitem{endnote_1}
K{\"{a}}stner and Voigtl{\"{a}}nder observed a similar phenomenon---the growth of Ge coverage caused the increase of elongated huts  density \cite {Kastner}.

\bibitem{atomic_structure}
L. V. Arapkina and V. A. Yuryev, 
JETP Lett. {\bf 91} 281 (2010).


\bibitem{endnote_3}
   A new terminology was specially introduced by us in Ref.~\onlinecite{classification} to emphasize that the difference between the clusters is not only in the shapes but primarily in their atomic structures. The wedge-like cluster may be short and nearly square-based but its atomic structure remains the structure of the wedge and not that of the pyramid \cite{classification,atomic_structure}. The term ``elongated'' implies elongation of some precursor. Usually the ``square-based'' clusters (the pyramids) are considered as such precursors. This assumption turned out to be wrong \cite{classification}. Note also that 
according to 
Ref.~\onlinecite{atomic_structure}  
transformation of a wedge-like hut cluster to a dome cluster is impossible too.



\bibitem{Goldfarb_JVST-A}
I. Goldfarb, J. H. G. Owen, D. R. Bowler, C. M. Goringe, P. T. Hayden, K. Mik, D. G. Pettifor, and G. A. D. Briggs,
J. Vac. Sci. Technol. A {\bf 16}, 1938 (1998).

\bibitem{Vailionis}
A. Vailionis, B. Cho, G. Glass, P. Desjardins, D. G. Cahill, and J. E. Greene, 
Phys. Rev. Lett. {\bf 85}, 3672 (2000).



\bibitem{sabelnik}
V. A. Yuryev, L. V. Arapkina, V. A. Chapnin, V. P. Kalinushkin, N. V. Kiryanova, O. V. Uvarov, K. V. Chizh,  R. O. Stepanov, L. A. Krylova, A. V. Voitsehovsky, and S. N. Nesmelov,
{\it Report on  ``Sabelnik-2'' Research Project} (Prokhorov Gen. Phys. Inst. RAS, Moscow, Russia, 2008) RF  State Reg. No.\,1603925;  
L. V. Arapkina,
V. A. Yuryev, K. V. Chizh, V. A. Chapnin,
{\it Proc. XIV Int. Symp. ``Nanophysics and nanoelectronics'', Nizhni Novgorod, Russia, 2010} (Inst. Microstruct. Phys. RAS, Nizhni Novgorod, Russia, 2010) vol. 2, p.~531.

\bibitem{epinucleation}
R. G. Pala and F. Liu, Phys. Rev. Lett. {\bf 95}, 136106 (2005).

\bibitem{hut_stability}
G.-H. Lu and F. Liu, Phys. Rev. Lett. {\bf 94}, 176103 (2005).

\bibitem{endnote_7}
It might seem that we studied nucleation of so-called ``pre-pyramids'' \cite{Vailionis} (or ``pre-huts''). We tried to observe the ``pre-huts'' carefully exploring arrays with $h_{\rm Ge}$ from 3 to 14~\r{A} but we failed. 
{\it The clusters always had their specific  structure}\cite{atomic_structure} which was independent of the width-to-height ratio and arose when the cluster second layer formed. 


\bibitem{our_Si(001)_en}
L. V. Arapkina,  V. M. Shevlyuga, and V. A. Yuryev, 
JETP Lett. {\bf 87}, 215 (2008).

\bibitem{endnote_4}
   Or more accurately the thickness of the Ge film measured by the graduated in advance film thickness monitor with the quartz sensor installed inside the MBE chamber.


\bibitem{WSxM}
I. Horcas, R.~Fernandez, J. M. Gomez-Rodriguez, J. Colchero, J. Gomez-Herrero, and A. M. Baro,
Rev. Sci. Instrum. {\bf 78}, 013705 (2007).

\bibitem{endnote_5}
The $c(4\times 2)$ reconstruction is also often revealed together with the $p(2\times 2)$ one \cite{atomic_structure} which confirms that both reconstructions lie  close in energy \cite{Ge_surface_energies}.

\bibitem{Ge_surface_energies} 
M. J. Beck, A. van de Walle, and M. Asta, 
Phys. Rev. B {\bf 70}, 205337 (2004).


\bibitem{fig1}
The same structure is observed on  top of a mature pyramid shown in Fig.~\ref{fig:example}(a).


\bibitem{ridge}
A complete set of ridge configurations arising during the wedge height growth can be easily obtained when drawing schematic plots of a growing wedge by serial---layer by layer from bottom to top---completion of its terraces.

\bibitem{facet_growth}
Of course, the triangular facets  also grow by the same number of layers as trapezoidal ones in the process of the in-height growth of a wedge, otherwise the whole crystalline structure of the cluster would be disturbed. As distinct from the in-height growth, rapid growth of the triangular facets, which is responsible for cluster elongation, does not cause the growth of the trapezoidal facets. 






\bibitem{thanks}
We thank one of the anonymous referees of this article for the interesting remark.




\end{thebibliography}
\end{document}